\documentclass[preprint,11pt]{aastex}

\begin{document}
\title{The Origins of Two Classes of Carbon-Enhanced, Metal-Poor Stars}

\author{Sean G. Ryan\altaffilmark{1}, Wako Aoki\altaffilmark{2}, John
E. Norris\altaffilmark{3}, Timothy C. Beers\altaffilmark{4}}

\altaffiltext{1}{Department of Physics and Astronomy, The Open
University, Walton Hall, Milton Keynes, MK7 6AA, UK; email:
s.g.ryan@open.ac.uk} 
\altaffiltext{2}{National Astronomical Observatory, Mitaka, Tokyo,
181-8588 Japan; email: aoki.wako@nao.ac.jp}
\altaffiltext{3}{Research School of Astronomy and Astrophysics, The
Australian National University, Mount Stromlo Observatory, Cotter
Road, Weston, ACT 2611, Australia; email: jen@mso.anu.edu.au}
\altaffiltext{4}{Department of Physics and Astronomy and JINA: Joint Institute
for
Nuclear Astrophysics, Michigan State
University, East Lansing, MI 48824-1116; email: beers@pa.msu.edu}

\begin{abstract} 


We have compiled composition, luminosity, and binarity
information for carbon-enhanced, metal-poor (CEMP) stars reported by
recent studies. We divided the CEMP star sample into two classes,
having high and low abundances, respectively, of the $s$-process
elements, and consider the abundances of several isotopes, in
particular $^{12}$C, $^{13}$C, and $^{14}$N, as well as the likely
evolutionary stages of each star. Despite the fact that objects in
both groups were selected from the same surveys (primarily the HK
survey), without {\it a-priori} knowledge of their $s$-process element
abundances, we identify the following remarkable difference between
the two classes: $s$-element-rich CEMP (CEMP-s) stars occupy a wide
range of evolutionary states, but do not have a strongly evolved
$^{13}$C/$^{14}$N ratio, whereas $s$-element-normal CEMP stars (CEMP-no) are
found only high up the first-ascent giant branch, and possess
$^{13}$C/$^{14}$N ratios approaching the CN-cycle equilibrium value.

We argue that these observational constraints can be accommodated by the
following scenarios. CEMP-s stars acquire their distinctive surface compositions
during their lifetimes when mass is transferred from an AGB companion that has
recently synthesised $^{12}$C and $s$-process elements. Such mass-accreting
stars can be enriched at almost any stage of their evolution, and hence will be
found throughout the HR diagram. Dilution of transferred surface material as the
accretor ascends the giant branch, and its surface convective zone deepens, may
reduce the number of such stars whose surfaces remain C-rich at high
luminosities. Many, but not necessarily all, such stars should currently be in
binary systems. Li-preserving CEMP-s stars may require a different explanation.
In contrast, a CEMP-no star is proposed to have formed from gas that was
enriched in $^{12}$C from the triple-alpha process in a previous generation of
stars, some of which has been converted to $^{13}$C and $^{14}$N during the
present star's giant-branch evolution. The binary fraction of such stars should
be the same as that of non-carbon-enhanced, metal-poor stars.

\end{abstract} 

\section{Introduction}

One unanticipated discovery of the HK-survey \citep{beers85,beers92,beers99}
search for extremely metal-poor stars was the realization that a large fraction
of metal-poor objects exhibited large carbon over-abundances, with [C/Fe]
reaching up to 2.0 dex or beyond (Rossi, Beers, \& Sneden 1999; Rossi et al.
2005). The carbon excesses are most obvious for stars with [Fe/H] $<$
$-2.5$.\footnote{[A/B] $\equiv \log(N_{\rm A}/N_{\rm B})- \log(N_{\rm A}/N_{\rm
B})_{\odot}$.} It is not yet clear whether the mechanism responsible for the
enrichment ceased, and only plays a small role at higher metallicity, or whether
it continued to operate at a constant absolute level which progressively became
insignificant in comparison to the increasing metallicity of the interstellar
medium as the young Galactic halo was chemically enriched. The decrease with
metallicity of the upper envelope of [C/Fe] for stars in the Rossi et al. (1999,
2005) studies is consistent with the latter possibility (Ryan 2003), but as the
source of the carbon is not yet known, we cannot be certain at present.
Estimates of the frequency of carbon-enhanced, metal-poor (CEMP) stars in
ongoing surveys for metal-poor stars ranges between 10\% and 25\%, depending on
how the sample is defined \citep{beers92,beers99,christlieb03}. Around 70\%
(Aoki et al. 2003) to 80\% (Tsangarides 2005) of the CEMP
stars appear also to be rich in $s$-process elements.

Carbon-rich stars belonging to the halo population, the so-called CH
stars (Keenan 1942), have been studied by a number of previous authors. These
stars have relatively low metallicity and large excesses of
neutron-capture elements associated with the $s$-process (Vanture
1992). The excesses of both carbon and the $s$-process elements
suggest a large contribution from AGB nucleosynthesis, although these
CH stars appear not to be in the AGB phase at present. A clear
explanation of their chemical nature was given by McClure and his
colleagues \citep{mcclure84,mcclure90} who found that most, if not
all, CH stars are members of binary systems.  This strongly suggested
that mass transfer to a surviving companion from an AGB star (which is
now evolved to a white dwarf) is their likely origin.  A similar
interpretation can be applied to the more metal-rich Ba stars
\citep[e.g., ][]{mcclure90}, as well as to subgiant CH stars
\citep[e.g., ][]{luck91,smith93}, indicating that this mechanism of
carbon enrichment likely operates in both metal-rich and metal-poor
binary systems.

One important question is whether this same scenario, based on AGB
nucleosynthesis and mass-transfer in binary systems, can fully explain
the CEMP stars found in recent surveys for metal-poor stars. In an
effort to uncover the origin(s) of the carbon excesses, we embarked on
a program of high-resolution spectroscopy using the Anglo-Australian
Telescope and the Subaru Telescope. Our intention was to measure
element abundance ratios that could be related to various
nucleosynthetic environments, and to obtain multi-epoch radial
velocity measurements that might identify the possible role of
binarity in the formation of these objects. It soon became clear,
based on our work, and that of other groups, that a range of
nucleosynthetic signatures existed amongst the CEMP stars. For
example, Sneden et al. (1996) identified a strong $r$-process
signature in CS~22892--052, while Norris et al. (1997a) and Barbuy et
al. (1997) found strong $s$-process signatures in four other CEMP
stars (two of which were originally discovered outside the HK survey
of Beers et al.), while Norris et al. (1997b) and Bonifacio et
al. (1998) found normal neutron-capture abundances in the CEMP star
CS~22957--027. Despite the excitement that surrounded the discovery of
the $r$-process-rich star CS~22892--052, no other CEMP stars with
excesses of r-process elements have been found. All remaining CEMP
stars from the HK survey observed to date at high spectroscopic
resolution either exhibit $s$-process signatures (CEMP-s) or normal
neutron-capture abundances (CEMP-no).

A recent summary of the diverse classes of CEMP stars can be found in
Aoki et al. (2002a, their Figure 8), where [Ba/Fe] acts as a
neutron-capture diagnostic, while [(C+N)/Fe] is a measure of the
number of C and N nuclei, since the abundances of these species may
have been re-distributed by the CN cycle. Aoki et al. proposed that
discrete classes of Ba-rich and Ba-normal stars can be
distinguished. In this paper we examine this proposed separation of
the CEMP stars into $s$-rich and $s$-normal classes in more detail,
and discuss scenarios for the formation and evolution of these objects
based on element and isotopic abundances measured in recently
published studies, as well as on the luminosity distributions of the
two classes.

\section{Sample Stars and Their Compositions}

CEMP stars for which measurements of $^{13}$C/$^{12}$C, N, which we
assume to be pure $^{14}$N, and Ba have been reported, are listed in
Table~1. Figure~1 shows their [Ba/Fe] and [Fe/H] values superposed on
the distribution of [Ba/Fe] for non-CEMP stars, using data compiled by
Norris, Ryan \& Beers (2001). For the purposes of this work, we define
C-rich to mean [C/Fe] $\ge$ 0.5, and Ba-rich to mean [Ba/Fe] $\ge$
0.5. Since the CEMP stars were observed without prior knowledge of
their neutron-capture-element abundances, the apparently clear
division separation into $s$-process-rich and $s$-process-normal stars
seen in Figure~1 is not due to any explicit selection criteria. It should
be noted that, even though studies of CEMP stars are rapidly
increasing, objects for which all of the abundances of C isotopes, N,
and Ba are determined are still quite limited. Hence, further detailed
abundance studies of such objects are strongly desired.

We suspect that further separation of the Ba-rich stars into
$s$-process and $r$-process stars, or some combination of the
processes, may also be relevant.  Most Ba-rich objects in Table 1 have
[Ba/Eu] larger than 0, indicating that Ba principally originates from
the $s$-process. There are a few CEMP stars that have much lower
[Ba/Eu] values than that of the $s$-process component in solar system
material: e.g., CS~29497--034 ([Ba/Eu]$=-0.22$: Hill et
al. 2000); HE~2148--1247 ([Ba/Eu]$=+0.38$: Cohen et al. 2003).
These examples suggest a significant contribution from the $r$-process
to Eu. (An alternative possibility is a contribution of the
$s$-process with quite different condition, e.g., with very high
neutron density; Johnson \& Bolte 2004). However, the [Ba/Eu] of these
objects is still much higher than that of the $r$-process
element-enhanced stars ([Ba/Eu]$\sim -0.8$; e.g., Sneden et al. 1996,
2003). Hence, a large contribution of the $s$-process to Ba should be
assumed. We presume for now that most or all of the Ba-rich stars in
our sample carry an $s$-process signature, and we use the expressions
Ba-rich and $s$-rich interchangeably. We note that the carbon- and
$r$-process-element-enhanced star CS~22892--052 is not included in our
sample, as it seems likely (given that this is the only star in which both
carbon and pure $r-$process enhancement have been detected to date)
that the carbon enhancement and $r-$process
enhancement may not be causally connected.

Figure~2a shows [C/N] as a function of $^{12}$C/$^{13}$C for our
sample of CEMP stars. Filled circles signify Ba-rich stars, while open
squares are Ba-normal stars. In this figure, the equilibrium values of
the CN-cycle (see Table~1) are shown by the {\it cross} at the lower
left. Inspection of this figure suggests that a correlation exists between
these two abundance ratios, as (A) the $^{12}$C/$^{13}$C ratio
decreases from $\sim$40 to $\sim$10 without large changes in [C/N],
and (B) the $^{12}$C/$^{13}$C ratio decreases from $\sim$10 to $\sim$3
with declining [C/N]. While sequence (B) indicates the CN-cycle
approaching equilibrium, sequence (A) suggests a partial CN-cycle in
which $^{13}$C burning is limited by the short timescale of the
process, or by a limited supply of protons (see, e.g., Aoki et
al. 2002a).

Whereas Figure~2(a) presents the commonly derived ratio [C/N], a more
physically significant ratio is $^{13}$C/$^{14}$N, since both of these
isotopes are produced only in the CN-cycle. Thus, any change in this
ratio must reflect CN-cycle activity, whereas [C/N] can be influenced
by pure $^{12}$C production in the triple-alpha
process\footnote{Strictly, N can also be produced in the ON-cycle, but
this is less important than the CN-cycle in low-mass stars, and in any
case is still an indication of shell-hydrogen burning.}. We assume
here that all of the N is $^{14}$N. The importance of considering
isotopes rather than just elements is illustrated by the Sun, whose
$^{13}$C/$^{14}$N = 0.04 is within a factor of 4 of the equilibrium
value (0.01), but whose $^{12}$C/$^{13}$C = 90 is a factor of 30 from
equilibrium (3). It is likely that the pre-solar nebula contained
$^{12}$C, $^{13}$C and $^{14}$N produced in a source (or sources) that
approached CN-cycle equilibrium, and substantially more pure $^{12}$C
from a triple-alpha source that elevated the $^{12}$C/$^{13}$C ratio
well above the CN-cycle value.

Figure~2(b) shows log$_{10}$($^{13}$C/$^{14}$N) vs
log$_{10}$($^{12}$C/$^{13}$C).  The sequences (A) and (B) discussed in
connection with [C/N] are, as expected, also seen in Figure~2(b). The
Sun no longer sits on the sequence. The pattern is broadly consistent
with the CN-cycle's approach to equilibrium, illustrated for one
particular model at 20$\times 10^6$~K by Clayton (1968, his
Figure~5-15), where the $^{12}$C/$^{13}$C ratio approaches equilibrium
long before $^{13}$C/$^{14}$N. We point out here that the advanced
stages of sequence (B) seem to be occupied only by the Ba-normal
stars. This provides an important first clue to the different origins
of the CEMP-s stars on the one hand, and the CEMP-no stars on the
other.

\section{The Distinct Origins of CEMP-s and CEMP-no Stars}

As may be seen in Figure~2, the CEMP-s stars all have
$^{13}$C/$^{14}$N ratios close to 1 ($-0.6$ $\le$
log$_{10}$($^{13}$C/$^{14}$N) $<$ 0.8), whereas the CEMP-no stars
extend over the full range of values extending down to the CN
equilibrium value. A second distinction between the two groups is
found in the H-R diagram. Using spectroscopically-determined surface
gravities, we calculate radii for the stars, assuming identical masses
of 0.85~M$_\odot$, and combine these with adopted effective
temperatures to calculate luminosities\footnote{$L \propto (M/g)T_{\rm
eff}^4$; uncertainties in these values should render $L$ accurate to
about 0.3 in the logarithm.}. The resulting H-R diagram is shown in
Figure~3. It is notable that the CEMP-no stars are found only at the
highest luminosities, log$_{10}$($L$/L$_\odot$) $^>_\sim$ 2.0, whereas
the CEMP-s stars are found throughout the post-main-sequence subgiant
and giant branches. It is also worth noting that, despite their strong
$s$-process signatures, the CEMP-s stars clearly are not currently AGB
stars, whose luminosities would exceed $\log_{10}(L$/L$_\odot$) $\sim
3-4$.

Combining these two pieces of information suggests a distinction
between the origins of these two subclasses of CEMP stars, which we
consider below.

\subsection{C-rich, Ba-normal stars}

Two of the CEMP-no stars (CS~22949--037 and CS~29498--043) also
exhibit over-abundances of oxygen and the $\alpha$ elements. These are
first-ascent giants that presumably began their lives with very large
$^{12}$C abundances. The low iron-peak-element content suggests that
enrichment of the gas benefited from mass loss from a massive stellar
envelope, or from a supernova which ejected little of the material
from its deepest layers (e.g., Tsujimoto \& Shigeyama 2003; Umeda \&
Nomoto 2003). This hypothesis requires low-mass star formation from
material that was not yet well-mixed in the very early Galaxy.

Here we examine the interpretation that the CEMP-no stars,
including the above two objects, formed from C-enriched gas.

\noindent 
(1)Binarity: If the CEMP-no stars are congenitally C-rich, then the
binary fraction of such stars should be the same as that of non-CEMP
stars.  Table~1 shows that little is currently known about the binary
status of this group of stars, but the data that exist presently do not
contradict this requirement.\footnote{One of the referees has suggested that the R
stars are the disk analogues of our CEMP-no stars, and that \citet{mcclure97}
has shown that none of these is a spectroscopic binary,
suggesting that they are the product of the coalescence of close
binaries which have mixed carbon to their surface following He-flash.
While we cannot dismiss this scenario for all members of the CEMP-no
class, we note that at least it does not apply in the case of the
CEMP-no binary CS~22957--027 in Table 1.  One must await more complete
radial velocity data to test the referee's suggestion more fully.}

\noindent
(2)Evolutionary stages: The CEMP-no stars are found only at the
top of the giant branch (Figure 3). This could be a sampling effect,
and CEMP-no stars near the main-sequence may remain to be
discovered. The fraction of stars near the main-sequence turn-off in
the full HK sample is about 50\% (Beers et al. 1992; Beers
1999). However, CEMP giants have lower temperatures than
main-sequence stars, and thus would be more easily identified by the
strength of their molecular carbon bands (such as the CH G-band) than for main-sequence
stars, if a similar level of carbon excess is assumed. The
distribution of CEMP-no stars in $T_{\rm eff}$ and in luminosity
shows a clear contrast to that of CEMP-s stars, which likely arose
due to mass accretion from an AGB companion (see next
subsection).

\noindent
(3)C and N abundances: CN-cycling was probably unimportant during
their main-sequence evolution, as these are low-mass stars. Their
surface chemical abundances would therefore not be modified at that
phase. However, during their post-main-sequence phase, the products of
the CN-cycle appear at the surface as a result of first dredge-up, in
much the same way as they do for conventional (non-CEMP) low-mass
stars. In fact, the evolution of the $^{12}$C/$^{13}$C ratio as a
function of luminosity (see Figure~4) appears consistent with that of
normal Pop.~II first-ascent giants in the globular cluster M4 and in the
field, and might be explained in terms of deeper mixing than standard models
predict (see Charbonnel 1995, her Figure 1) with $^{12}$C/$^{13}$C
ratios of 3-10 for $\log_{10}$($L$/L$_\odot$) in the range 2-3.

If CS~22949--037 is to be viewed in the context of this model, how
should we view its high N abundance? Norris et al. (2001, 2002)
suggested this may originate in N-producing hypernovae, whereas
Depagne et al. (2002) associated it with CN-processing subsequent to a
low-energy supernova. The $^{13}$C/$^{14}$N ratio in CS~22949--037 (see
Figure~2(b)) is certainly consistent with CN-cycling having been
active, as suggested by Depagne et al., but whether this occurred
after a low-energy supernova that might account for the low iron yield
(relative to magnesium and carbon) or after a hypernova that produced
a high nitrogen yield directly, is not yet settled.

A possible argument {\it against} CS~22949--037 being a member of the
proposed class of CEMP-no stars is its position in the [C/N]
vs $T_{\rm eff}$ diagram (Figure 12 of Aoki et al. 2002b). Here the
star is seen at lower [C/N] than other stars at its temperature, as if
it does not belong to the same evolutionary sequence. This may
indicate that the high nitrogen abundance of CS~22949--037 is not due
to internal processes within the star, but rather may originate from a
massive progenitor, as pointed out by Aoki et al. (2002b). An
alternative possibility is that internal mixing was more effective in
CS~22949--037 than in other stars with similar effective
temperatures. The efficiency of internal processes may not only
be a function of the effective temperature, but could also be related to
stellar rotation, metallicity etc., and might not easily be predicted
for individual objects.

The above facts at least do not contradict our hypothesis that
CEMP-no stars formed from C-rich gas, although explicit support of this
scenario is still weak, except for the two O and $\alpha$
element-enhanced stars. An alternative explanation is that the
abundance peculiarities of such stars occurs at the top of the red
giant branch in a mechanism quite different to the congenital one
proposed here. Fujimoto et al.  (2000) have calculated that in
low-mass, very metal-poor giants, convection will carry C that has
been freshly synthesised in the He flash up to the H-rich layers and
eventually into the surface convective zone, where $^{12}$C-, $^{13}$C-
and $^{14}$N-rich metal-poor post-red-giant branch stars will be
produced. It is not clear whether the luminosities of our stars are
appropriate to that phase of evolution, but chemically at least the
CEMP-no stars are similar to those envisaged by Fujimoto et al (2000).

\subsection{C-rich, Ba-rich stars}

CEMP-s stars are found to encompass a wide range of evolutionary states, from the
main-sequence turnoff to the tip of the red-giant branch. They exhibit high
$s$-process abundances and, although $^{13}$C/$^{12}$C approaches the equilibrium
ratio, $^{13}$C/$^{14}$N does not, although it too is lower in the higher
luminosity stars of this type. These factors, combined with the high $s$-process
abundances, suggest that most, if not all, of their chemical anomalies are extrinsic
to these stars.

It has been proposed that most members of this group of stars formed
from ``normal'' ISM material, but then later accreted C- and $s$-rich gas as CH
stars, Ba stars and subgiant CH stars
\cite[e.g.,][]{mcclure80,mcclure84,smith80, han95}. This gas
apparently had undergone some hydrogen burning to produce $^{13}$C,
but either on short timescales, or in proton-poor environments that
prevented complete evolution of the $^{13}$C/$^{14}$N ratio toward
equilibrium. Comparison of these stars' $^{13}$C/$^{14}$N abundances
with the time-tags in Clayton's calculation (see our Figure 2(b))
would suggest a timescale less than 3$\times 10^3$~yr. One such
environment where this might occur is the He-intershell of a
thermally-pulsing AGB star, which is proton-poor and undergoes
nucleosynthesis on much shorter timescales than is required for the
$^{13}$C/$^{14}$N ratio to achieve equilibrium. This environment is
also thought to be the most important site for production of the
$s$-process elements.

If the element anomalies were purely inherited from a companion, then the
isotope ratios would be independent of the evolutionary state of the accretor.
Figure~4 shows that even the low-luminosity Ba-rich stars have low
$^{12}$C/$^{13}$C ratios, quite unlike normal halo giants. There is in fact
little, if any, change of this ratio with position in the H-R diagram.

We noted above that the luminosity distributions of the CEMP-s and
CEMP-no stars differ considerably.\footnote{LP~706--7 and LP~625--44
were first recognized as C-rich in proper-motion-selected studies, and
so it might be argued that they therefore inflate the numbers at low
luminosity. However, LP~706--7 was independently re-discovered in the
magnitude-limited HK survey that generated most of the other objects
in this study, so this star would stay in the sample even if we
removed proper-motion-selected objects. That is, LP~625--44 alone might
be removed on this basis.} The fact that the CEMP-s stars are not
primarily high-luminosity objects may be additional evidence that their
abundance properties were accreted. Stars that formed from gas with
peculiar abundances will possess that material {\it throughout their
structure}, whereas accretors will have abnormal abundances in 
their {\it surface zones} exclusively. As
the latter objects ascend the giant branch, the surface material will
be diluted by normal material as the convective zone deepens.  Hence,
one might expect accretors to drop out of the sample of C-rich stars
as they ascend the giant branch. This may explain the greater fraction
of low-luminosity CEMP-s objects compared with the existence
of only high-luminosity CEMP-no stars in the complete sample of C-rich
objects provided (essentially) from the HK survey. If accreted systems
do indeed dilute their envelopes as they evolve, the survival of at
least some of these objects to high luminosities must reflect the
large fractions of the envelope mass accreted.

This discussion seems to be the opposite of the suggestion by
\citet{luck91} that the excesses of carbon and $s$-process elements
found in subgiant CH stars are not diluted by the deepening of the
convective envelope when the star becomes a red giant (a Ba star), and
that subgiant CH stars have thick mantles of material accreted from
companion AGB stars. However, the average metallicity of CEMP stars
discussed here is more than {\it two dex lower} than that of subgiant CH
stars and Ba stars. The effect of mass transfer from a companion AGB
star should appear more clearly in metal-poor stars, and objects in
which only a small amount of material was accreted might therefore be included 
amongst our unevolved CEMP stars.

A possible weakness with this proposal for the CEMP-s stars is that
several of them are known to exhibit no detected radial-velocity
variation over long periods (Norris et al. 1997a; Preston \& Sneden
2001), as shown in Table 1. Such stars provide no evidence of a
companion that could previously have deposited the $s$-process
material onto its surface.  However, a Monte Carlo analysis of 
the observed radial velocity data for a sample of 19 CEMP-s stars led 
Lucatello et al. (2005) to conclude that an observed binary fraction of 68\% was
consistent with {\it all} CEMP-s stars being in multiple systems. The detection
efficiency of such systems was found to be significantly below 100\%, because of
the finite timespan of extant radial velocity data, the possibly wide range of
periods in the parent distribution, inclination effects, and the finite accuracy
of the measurements, amongst other factors. Alternatively, Preston \& Sneden
(2001) suggested a path to the carbon-enhanced subgiant stars with no binary
interaction to explain these apparently single CEMP-s stars. 

More troublesome for the mass-transfer scenario are the stars LP~706--7
and CS~22898--027, which exhibit high Li/H ratios, consistent with the
primordial value (Thorburn 1994; Aoki et al., in preparation). The
latter constraint requires either that the matter transferred from an
unidentified companion also had the primordial Li/H ratio, which is
most unlikely for an AGB star whose envelope has been mixed with deep
layers where He-burning at 10$^8$ K and the $s$-process has occurred,
or the star was enriched by almost pure C and $s$-process material
containing little H to dilute the primordial Li/H value. Complete
mergers of a binary pair to form the single star seen today would not
preserve the primordial Li/H ratio either, so a non-binary origin must
be found. Nor can we appeal to non-standard, internal mixing scenarios
to regenerate a star that has already gone through He-burning to
produce excess C, because such objects, if they exist at all, would
not preserve the primordial Li/H value. The unavoidable conclusion
seems to be that LP~706--7, CS~22898--027, and possibly other apparently
single, Li-preserving CEMP-s stars did not form by the accretion
mechanism, and need not be binaries.  Of the four apparently single
CEMP-s stars, two are known to be Li-bearing. It may be that there are
two distinct classes of CEMP-s stars: Li-bearing single stars, and
Li-poor, mass-transfer binaries.

\section{Concluding remarks}

We have shown that most CEMP stars divide cleanly into Ba-rich and Ba-normal
objects, which we refer to as CEMP-s and CEMP-no stars, respectively.

The isotopic compositions of the two classes and their different luminosity
distributions have led us to propose that the two classes have different origins. We
associate the CEMP-no sample with stars that formed with abnormally high
$^{12}$C abundances, which they inherited from the gas clouds from which they
formed. Possible mechanisms by which the gas clouds were enriched include
supernovae whose mass cut was exterior to most heavier metals, or mass-losing
massive stars. These congenitally C-rich objects have processed the $^{12}$C
through the CNO-cycle during their shell-hydrogen-burning first-ascent
red-giant-branch phase, and have dredged $^{13}$C and $^{14}$N to the surface in
much the same way as normal low-mass stars. We speculate that such stars are
found preferentially near the upper end of the red-giant branch with low
temperatures, because the enhancement of carbon is identified by the strength of
absorption bands of carbon-bearing molecules. In contrast, the CEMP-s sample
is found over a wide range of evolutionary stages, and while it possesses a
similar $^{12}$C/$^{13}$C distribution to the $s$-normal stars, $^{14}$N is not
as heavily processed, which we believe may indicate shorter nucleosynthetic
timescales and/or proton starvation of the CN-cycle. We propose that binary
stars in this group accreted material enriched in $^{12}$C, $^{13}$C, a little
N, and $s$-process elements from a now-extinct AGB companion. As these stars
climb the first-ascent giant branch, their accreted layers will be diluted with
original envelope material, and in some cases the stars will drop out of the
CEMP-s class. This could explain why high-luminosity (low-temperature)
members do not completely dominate this class. Non-binary members of the CEMP-s
group, on the other hand, may have formed from gas enriched by such objects,
i.e., be congenitally abnormal, especially if they retain significant Li
abundances, as two out of the four possibly-single CEMP-s do.

We believe that the full diversity of CEMP stars has yet to be
discovered observationally, let alone explained theoretically. Hence,
continued characterisation and analysis of these objects is
surely warranted. Some stars, such as LP~706--7, still present challenges to
the AGB-accretion picture; we expect additional examples of
non-conforming stars to be identified. Nevertheless, we are of the
view that the distinctions set out in this work, while still
incomplete, may shed light on the diverse origins of the CEMP stars.

\acknowledgments

We thank Dr. G. W. Preston for his useful comments as a referee. This
work was supported financially by PPARC (PPA/O/S/1998/00658). J.E.N.
acknowledges support from grant DP0342613 of the Australian Research
Council.  T.C.B. acknowledges partial support from grants AST
00-98508, AST 00-98549, and AST 04-06784, as well as from grant PHY
02-16783; Physics Frontier Center/Joint Institute for Nuclear
Astrophysics (JINA), awarded by the U.S. National Science Foundation.

\begin{deluxetable}{lccccccccccll}
\rotate
\tabletypesize{\footnotesize}
\tablewidth{0pt}
\tablecaption{Literature Data for C-rich Stars \label{tab1}}
\startdata
\tableline
\tableline
Star		&$T_{\rm eff}$&log$g$&log($L$/L$_\odot$)&[Fe/H]&[C/Fe]&[N/Fe]&$^{12}$C/$^{13}$C&$^{13}$C/$^{14}$N&[Ba/Fe]&[Eu/Fe]&binarity& References\\
\tableline
\multicolumn{11}{l}{\bf Comparison data}\\
Sun		&5777&4.44&0.00&0.00  &0.00&0.00   &   90&   0.04&   0.00& 0.00 & & \\
CN equilibrium	&    &    &    &      &    &       &    3&   0.01&       &      & & 1 \\
\multicolumn{11}{l}{\bf Ba-normal stars}\\
CS~22877--001	&5100&2.20&1.95&$-$2.17&1.00&0.00&$>$10&$<$3.46&$-$0.49 & $<0.6$  & ?   & 2 \\
CS~22949--037	&4900&1.70&2.38&$-$3.79&1.05&2.30&4&0.04&$-$0.84        & $<0.93$ & No? & 3,4 \\
CS~22957--027	&5100&1.40&2.75&$-$3.11&2.40&1.60&8&2.67&$-$1.23        & \nodata & Yes & 5,6\\
CS~29498--043	&4600&1.20&2.77&$-$3.54&2.09&2.27&6&0.22&$-$0.46        & \nodata & No? & 5,7\\
CS~29502--092	&5000&2.10&2.02&$-$2.76&1.00&0.70&20&0.36&$-$0.82       & $<0.4$  & ?   & 2\\
CS~30314--067	&4400&1.00&2.89&$-$2.85&0.50&1.20&5&0.13&$-$0.57        & $<-0.5$ & ?   & 2 \\
\multicolumn{11}{l}{\bf Ba-rich stars}\\
CS~22880--074 &     5850&     3.80&     0.59&   $-1.93$&     1.30&   $ -0.10$&  $>$40&  $<$2.33&     1.31& 0.5  & No? & 5,6 \\
CS~22881--036 &     6200&     4.00&     0.50&   $-2.06$&     1.96&     1.00&       40&     0.83&     1.93& 1.00 & No? & 6 \\
CS~22898--027 &     6250&     3.70&     0.80&   $-2.25$&     2.20&     0.90&       15&     4.74&     2.23& 1.88 & No?$^a$ & 5,6 \\
CS~22942--019 &     5000&     2.40&     1.72&   $-2.64$&     2.00&     0.80&        8&     6.70&     1.92& 0.79 & Yes & 5,6 \\
CS~22948--027 &     4600&     1.00&     2.97&   $-2.57$&     2.00&     1.80&       10&     0.55&     1.85& 1.57 & Yes & 2,8 \\
CS~29497--034 &     4800&      1.8&     2.25&   $-2.90$&     2.25&     2.30&       12&     0.26&     2.03& 2.25 & ?   & 8 \\
CS~30301--015 &     4750&      0.8&     3.23&   $-2.64$&     1.60&     1.70&        6&     0.43&     1.45& 0.2: & ?   & 5 \\
CS~31062--050 &     5600&     3.00&     1.31&   $-2.32$&     2.00&     1.20&        8&     2.67&     2.30& 1.84 & Yes & 5,9 \\
HD~196944    &     5250&     1.80&     2.40&   $-2.25$&     1.20&     1.30&        5&     0.50&     1.10& 0.17 & Yes & 5,10 \\
LP~706--7$^b$ &     6250&     4.50&     0.00&   $-2.55$&     2.10&     1.20&       15&     1.89&     1.98& 1.62 & No?$^a$ & 5,11 \\
LP~625--44    &     5500&     2.80&     1.48&   $-2.68$&     1.95&     1.65&       20&     0.36&     2.74& 1.97 & Yes & 12,13 \\
HE~2148--1247 &     6380&     3.90&     0.64&   $-2.32$&     1.91&     1.65&       10&     0.63&     2.36& 1.98 & Yes & 14 \\
HE~0024--2523 &     6625&     4.30&     0.31&   $-2.72$&     2.60&     2.10&        6&     1.72&     1.46& $<1.1$ & Yes & 15 \\
\tableline
\multicolumn{12}{l}{a: CS~22898--027 and LP~706--7 are Li-bearing stars as well as not having obvious companions.  }\cr
\multicolumn{12}{l}{b: LP 706--7 = CS~31062--012}\cr

References:&\multicolumn{11}{l}{(1) Arnould et al. 1999; (2) Aoki et al. 2002a; (3) Norris et al. 2002; 
(4) Depagne et al. 2002; }\\
	&\multicolumn{11}{l}{(5) Aoki et al. 2002b; (6) Preston \& Sneden 2001; (7) Aoki et al. (2004);
(8) Hill et al. 2000; } \\ 
        &\multicolumn{11}{l}{(9) Aoki et al. (2003); (10) Tsangarides (2005); (11) Norris et al. (1997a); }\\
        &\multicolumn{11}{l}{(12) Aoki et al. (2000); (13) Aoki et al. (2001); (14) Cohen et al. (2003); (15) Lucatello et al. (2003)
}\\
\tableline
\enddata
\end{deluxetable}

\clearpage

\begin{figure}
\includegraphics[width=14cm]{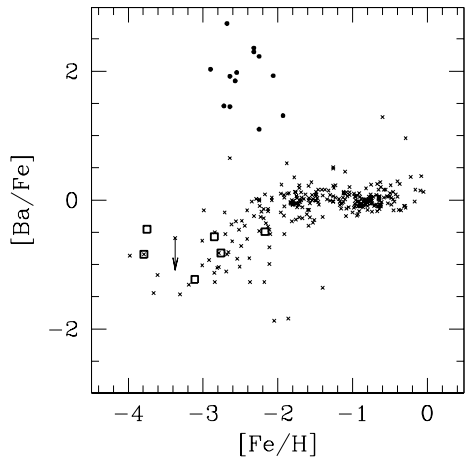}
\caption{
[Ba/Fe] vs [Fe/H] for the CEMP stars in Table~1, 
shown against a background of (apparently) C-normal stars. 
{\it Filled circles}: Ba-rich, C-rich (CEMP-s) stars; 
{\it open squares}: Ba-normal, C-rich (CEMP-no) stars;
{\it crosses}: stars with normal or unspecified C abundances.
}
\end{figure}

\begin{figure}
\includegraphics[width=14cm]{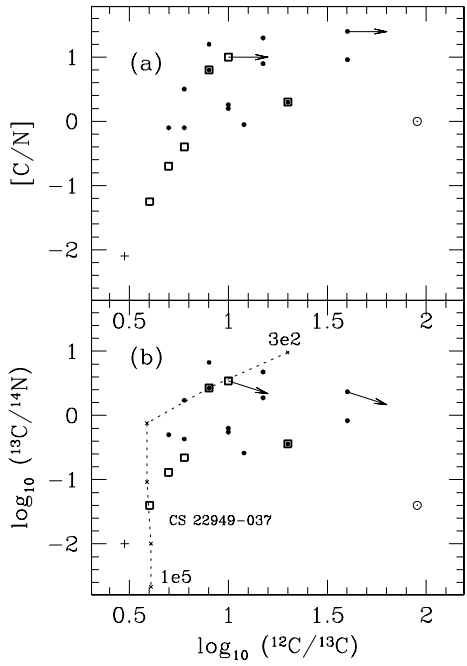}
\caption{
(a) [C/N] as a function of log$_{10}$($^{12}$C/$^{13}$C) for CEMP stars. 
{\it Filled circles}: Ba-rich, C-rich (CEMP-s) stars; {\it open squares}:
Ba-normal, C-rich (CEMP-no) stars.
The {\it cross} indicates the equilibrium values of the CN-cycle,
and the {\it Sun symbol} gives the solar locus.
(b) 
The same stars as in (a), but showing the CN-cycle-specific
isotopes log$_{10}$($^{13}$C/$^{14}$N). 
The dotted curve connects isotope calculations at 
$3\times 10^2$, 
$1\times 10^3$, 
$3\times 10^3$, 
$1\times 10^4$, 
$3\times 10^4$ and
$1\times 10^5$~yr for a 
CNO-cycle running at $20\times 10^6$~K by Clayton (1968).
}
\end{figure}

\begin{figure}
\includegraphics[width=14cm]{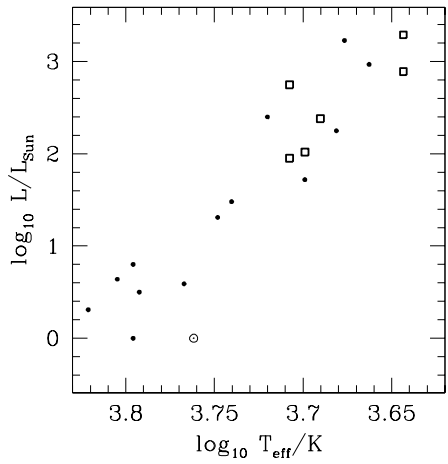}
\caption{
H-R diagram 
based on spectroscopic gravities and assumed masses 0.85~M$_\odot$
(except for the Sun).
{\it filled symbols}  Ba-rich, C-rich (CEMP-s) stars; {\it open symbols}
Ba-normal, C-rich (CEMP-no) stars. 
}
\end{figure}

\begin{figure}
\includegraphics[width=14cm]{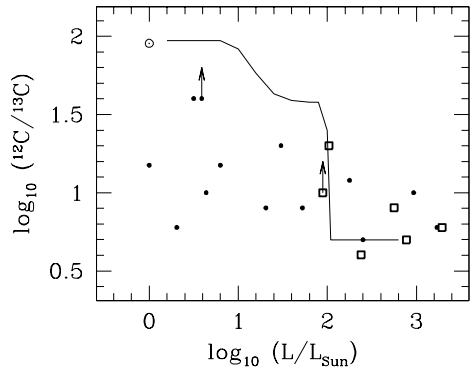}
\caption{Carbon isotope ratios as a function of luminosity.
{\it filled symbols}  Ba-rich, C-rich (CEMP-s) stars; {\it open symbols}
Ba-normal, C-rich (CEMP-no) stars. 
The solid curve is the evolutionary path for the extra-mixing scenario
of Charbonnel (1995), consistent with field halo stars and M4.}
\end{figure}

\end{document}